\documentclass[%
superscriptaddress,
 amsmath,amssymb,
 twocolumn,
 aps,
]{revtex4-2}

\usepackage{graphicx}% Include figure files
\usepackage{dcolumn}% Align table columns on decimal point
\usepackage{bm}% bold math
\usepackage[colorinlistoftodos]{todonotes}
\usepackage{xcolor}
\begin{document}

\title{Signatures of pressure-enhanced helimagnetic order in van der Waals multiferroic NiI$_2$}
%\title{Signatures of pressure-enhanced type-II multiferroic order in van der Waals helimagnetic NiI$_2$}

\affiliation{Department of Physics, Massachusetts Institute of Technology, Cambridge, MA 02139, USA}
\affiliation{HPCAT, Advanced Photon Source, Argonne National Laboratory, Lemont, IL 60439, USA}
\affiliation{Department of Physics, Arizona State University, Tempe, AZ 85287, USA}
\affiliation{Nanomat/Q-mat/CESAM, Universit\'{e} de Li\`{e}ge, B-4000 Sart Tilman, Belgium}
\affiliation{Consiglio Nazionale delle Ricerche CNR-SPIN, Area della Ricerca di Tor Vergata, Via del Fosso del Cavaliere 100, I-00133 Rome, Italy}
\affiliation{European Theoretical Spectroscopy Facility www.etsf.eu}
\affiliation{Department of Electrical Engineering and Computer Science, Massachusetts Institute of Technology, Cambridge, MA 02139, USA}
\affiliation{These authors contributed equally to this work}

\author{Connor A. Occhialini}
\email{caocchia@mit.edu}
\affiliation{Department of Physics, Massachusetts Institute of Technology, Cambridge, MA 02139, USA}
\affiliation{These authors contributed equally to this work}

\author{Luiz G. P. Martins}
\email{lmartins@alum.mit.edu}
\affiliation{Department of Physics, Massachusetts Institute of Technology, Cambridge, MA 02139, USA}
\affiliation{These authors contributed equally to this work}

\author{Qian Song}
\affiliation{Department of Physics, Massachusetts Institute of Technology, Cambridge, MA 02139, USA}

\author{Jesse S. Smith}
\affiliation{HPCAT, Advanced Photon Source, Argonne National Laboratory, Lemont, IL 60439, USA}

\author{Jesse Kapeghian}
\affiliation{Department of Physics, Arizona State University, Tempe, AZ 85287, USA}

\author{Danila Amoroso}
\affiliation{Nanomat/Q-mat/CESAM, Universit\'{e} de Li\`{e}ge, B-4000 Sart Tilman, Belgium}

\author{Joshua J. Sanchez}
\affiliation{Department of Physics, Massachusetts Institute of Technology, Cambridge, MA 02139, USA}

\author{Paolo Barone}
%\email{paolo.barone@spin.cnr.it}
\affiliation{Consiglio Nazionale delle Ricerche CNR-SPIN, Area della Ricerca di Tor Vergata, Via del Fosso del Cavaliere 100, I-00133 Rome, Italy}

\author{Bertrand Dup\'e}
\affiliation{Nanomat/Q-mat/CESAM, Universit\'{e} de Li\`{e}ge, B-4000 Sart Tilman, Belgium}

\author{Matthieu J. Verstraete}
\affiliation{Nanomat/Q-mat/CESAM, Universit\'{e} de Li\`{e}ge, B-4000 Sart Tilman, Belgium}
\affiliation{European Theoretical Spectroscopy Facility www.etsf.eu}

\author{Jing Kong}
\affiliation{Department of Electrical Engineering and Computer Science, Massachusetts Institute of Technology, Cambridge, MA 02139, USA}

\author{Antia S. Botana}
\affiliation{Department of Physics, Arizona State University, Tempe, AZ 85287, USA}

\author{Riccardo Comin}
\email{rcomin@mit.edu}
\affiliation{Department of Physics, Massachusetts Institute of Technology, Cambridge, MA 02139, USA}

\date{\today}

\begin{abstract} 

The van der Waals (vdW) type-II multiferroic NiI$_2$ has emerged as a candidate for exploring non-collinear magnetism and magnetoelectric effects in the 2D limit. Frustrated intralayer exchange interactions on a triangular lattice result in a helimagnetic ground state, with spin-induced improper ferroelectricity stabilized by the interlayer interactions. Here we investigate the magnetic and structural phase transitions in bulk NiI$_2$, using high-pressure Raman spectroscopy, optical linear dichroism, and x-ray diffraction. We obtain evidence for a significant pressure enhancement of the antiferromagnetic and helimagnetic transition temperatures, at rates of $\sim15.3/14.4$ K/GPa, respectively. These enhancements are attributed to a cooperative effect of pressure-enhanced interlayer and third-nearest-neighbor intralayer exchange. These results reveal a general path for obtaining high-temperature type-II multiferroicity via high pressures in vdW materials.

\end{abstract}

%here is a 600 character version of the abstract if needed
%The van der Waals (vdW) helimagnet NiI2 has emerged as a candidate for exploring multiferroic order in the 2D limit. Here, we investigate the magnetic and structural transitions in NiI2 using high-pressure Raman spectroscopy, linear dichroism and x-ray diffraction. We reveal a significant enhancement of the antiferromagnetic/helimagnetic transitions at rates of 15.3/14.4 K/GPa, respectively. We associate this behavior to pressure-enhanced interlayer and third neighbor intralayer exchange interactions, revealing a general path to stabilize multiferroic order via high pressures in vdW materials.

\maketitle

Transition metal dihalides are an emerging class of 2D van der Waals (vdW) materials presenting multiferroic order and non-collinear spin textures \cite{Song2022,McGuire2017,Bikaljevic2021,Botana2019,Amoroso2020,Liu2020,Kurumaji2013,Lebedev2023,Ju2021, Zhang2020b}. Type-II multiferroics with strongly coupled magnetoelectric effects have the potential to be integral components in functional vdW-based nanoscale devices \cite{Burch2018, Soler-Delgado2022, Lebedev2023, Kurumaji2013, Guan2020, Spaldin2019, Tokura2014,Jiang2018}. The magnetic semiconductor NiI$_2$ is a promising candidate to realize these goals, hosting a type-II multiferroic phase from the bulk to the single-layer limit \cite{Song2022, Lebedev2023, Kurumaji2013, Ju2021}. Identifying the microscopic mechanisms that stabilize the multiferroic state, and finding general routes to enhance the transition temperatures, are important steps towards realizing functional applications and finding new material candidates.

\begin{figure}[h]
\centering

\includegraphics[width = \columnwidth]{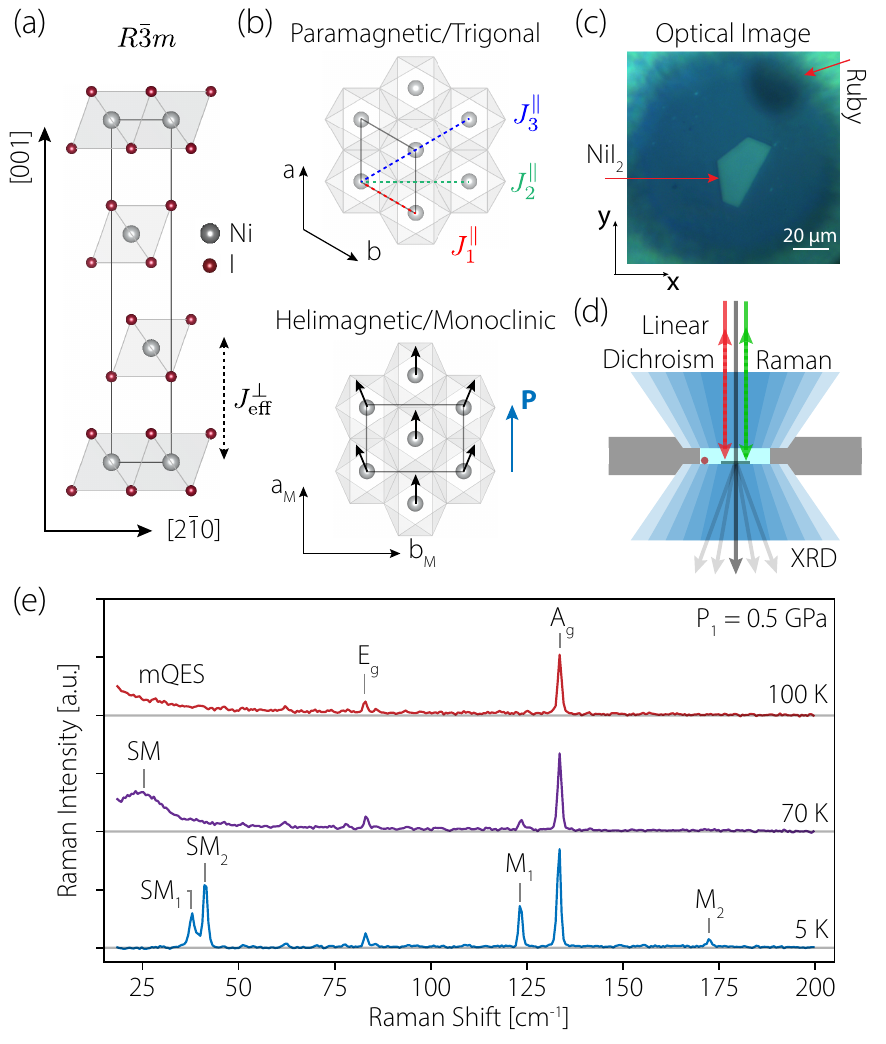}

\caption{(a), $R\bar{3}m$ structure of NiI$_2$ indicating the effective interlayer exchange, $J_\text{eff}^\perp$. (b) High-temperature triangular lattice indicating intralayer exchange pathways $J_n^\parallel$ for $n\leq 3$ (upper) and the monoclinic lattice below $T_{N,2}$ with helimagnetic spin texture (black arrows) and spin-induced polarization ($\mathbf{P}$, blue, lower). (c), Optical image of single-crystal NiI$_2$ sample inside the DAC used for optical experiments. Scale bar: 20 $\mu$m. (d), Schematic of DAC geometry for reflection LD/Raman measurements and (separate) transmission powder XRD measurements. (e), Representative Raman spectra at $P_1 = 0.5$ GPa and (i) 100 K ($T > T_{N,1}$), (ii) 70 K ($T_{N,2} < T < T_{N,1}$) and (iii) 5 K ($T < T_{N,2}$). The phonon modes ($E_g$/$A_g$) and the modes associated to magnetic order (mQES, SM, SM${}_{1/2}$, and M${}_{1/2}$) are indicated. Curves are offset for clarity with zero baselines indicated.}

\label{fig:fig1}
\end{figure}

NiI$_2$ forms a triangular lattice of magnetic Ni ions in the trigonal $R\bar{3}m$ structure (Fig. \ref{fig:fig1}(a)) \cite{Kuindersma1981, Kurumaji2013}. In the bulk, two magnetic phase transitions occur with decreasing temperature: first to a collinear antiferromagnetic (AFM) state at $T_{N,1} = 75$ K and then to a single-{\bf Q} proper-screw helimagnetic state below $T_{N,2} = 60$ K with wavevector $\mathbf{Q} = 0.138\mathbf{a^*} + 1.457 \mathbf{c^*}$ \cite{Kuindersma1981, Song2022}. The helimagnetism arises predominately from frustration between nearest neighbor ferromagnetic (FM, $J_1^\parallel$) and third-nearest neighbor AFM ($J_3^\parallel$) exchange within the triangular lattice planes (Fig. \ref{fig:fig1}(b)) \cite{Amoroso2020, Song2022, Kuindersma1981}. The transition to a single-{\bf Q} magnetic state is accompanied by a structural phase transition \cite{Kuindersma1981}, reducing the crystallographic symmetry from trigonal to monoclinic (space group $C_2$, (Fig. \ref{fig:fig1}(b))) \cite{Kuindersma1981, Kurumaji2013, Song2022, Ju2021}. Furthermore, the non-collinear magnetic state below $T_{N,2}$ hosts a spin-induced ferroelectric polarization that is tunable with magnetic field \cite{Kurumaji2013}, making NiI$_2$ a type-II multiferroic. Notably, this multiferroic phase persists to the monolayer limit with a reduced transition temperature of $T = 20$ K \cite{Song2022, Lebedev2023, Amoroso2020}. The continuous decrease of $T_{N,2}$ with layer number is attributed to a dimensionality-controlled reduction of the interlayer exchange interaction, $J_\text{eff}^\perp$ (Fig. \ref{fig:fig1}(a)) \cite{Song2022}. This finding immediately suggests that enhancing the interlayer interactions in NiI$_2$ could help stabilize the multiferroic phase at higher temperatures. 

In this vein, hydrostatic pressure is a powerful tuning knob for electronic and magnetic properties, particularly for vdW materials given the high sensitivity of interlayer interactions to interlayer distance \cite{Song2019,Valenta2021,Li2019,Yankowitz2018,Harms2022,Zhang2020b,Terada2022, Martins2023}. Several high-pressure studies on NiI$_2$ revealed an insulator-to-metal transition (IMT) around 16-19 GPa which is driven by a closure of the I-Ni ligand-to-metal charge transfer gap \cite{Chen1993,Dufek1995,DaSilva1992}. High-pressure M\"{o}ssbauer spectroscopy experiments revealed the presence of local magnetic order with an enhanced transition temperature up to 300 K before suddenly quenching near the metallization point \cite{Pasternak1990}. While this provides strong evidence for pressure-stabilized magnetic phase transitions, a distinction between the AFM and multiferroic orders had not been established at the time those studies were conducted. Thus, the possible stabilization of the helimagnetic order, and the associated multiferroic state, with hydrostatic pressure remains unknown.

Optical spectroscopy has proven to be a powerful method for interrogating both the AFM and the helimagnetic transitions in NiI$_2$ \cite{Liu2020,Song2022}. Several new modes appear in the Raman spectra below both $T_{N,1}$ and $T_{N,2}$ \cite{Liu2020,Song2022}. This includes a characteristic pair of low-energy modes connected to the helimagnetic transition, and attributed to the fundamental electromagnon excitations of the multiferroic state \cite{Song2022,Rovillain2010,Pimenov2006}. In addition, the multiferroic transition is characterized by significant breaking of lattice symmetries, including the loss of both three-fold rotational symmetry and inversion symmetry, induced by the polar single-{\bf Q} helical spin order \cite{Song2022,Fabrykiewicz2021,Kurumaji2013,Kuindersma1981}. These broken symmetries can be detected through optical linear dichroism and second harmonic generation signals, respectively, providing a robust optical signature of the multiferroic state \cite{Song2022,Ju2021,Kumdersma1981}. 

Here, we track the magnetic phase transitions in NiI$_2$ as a function of pressure using a combination of Raman spectroscopy, optical linear dichroism and powder X-ray diffraction. The optical response signals a large enhancement of both the $T_{N,1}$ and $T_{N,2}$ transitions, at rates of $\sim 15.3$ K/GPa and $14.4$ K/GPa, respectively, up to 5 GPa. Meanwhile, the magnetic phase transitions induce multiple crystallographic signatures, including a reduction of lattice symmetry to monoclinic, and an associated interlayer ($c$-axis) compression, confirming the magneto-structural transitions associated to helimagnetic order up to 10 GPa. The experimental trend for the magnetic transition temperatures are in quantitative agreement with calculated values determined from the pressure-dependent magnetic exchange interactions which have been recently reported in Ref.~\cite{theoryref}. This reveals a cooperative effect between the pressure-enhanced interlayer ($J^\perp_\text{eff}$) and intralayer ($J^\parallel_3$) exchange interactions, which stabilize the helimagnetic phase at high pressures.

We first discuss the pressure- and temperature-dependent optical spectroscopic experiments. We used a bulk-like (thickness $\sim 100$ nm) single crystal of dimension $\sim$ 40 x 20 $\mu$m grown by physical vapor deposition as previously reported \cite{Liu2020,Song2022}. The sample was loaded into a diamond anvil cell (DAC, Almax CryoDAC-ST) using 4:1 methanol-ethanol as the pressure transmitting medium. An optical image of the sample is shown in Fig. \ref{fig:fig1}(c) along with a schematic of the experimental geometry in Fig. \ref{fig:fig1}(d). Representative cross-polarized (XY) Raman spectra at $P_1 = 0.5$ GPa across both the $T_{N,1}$ and $T_{N,2}$ transitions are shown in Fig. \ref{fig:fig1}(e). Besides the two Raman-active phonon modes ($E_g$/$A_g$), the new features associated to the magnetic state include a magnetic quasi-elastic scattering (mQES) mode above $T_{N,1}$, which evolves into a finite-frequency soft mode (SM) below $T_{N,1}$ and further splits to two distinct modes SM${}_1$/SM${}_2$ below $T_{N,2}$. These modes have been attributed to electro-magnons of the multiferroic state \cite{Song2022}. Additionally, higher-energy magnon modes near $\sim 120$ cm${}^{-1}$ (M${}_1$) and 165 cm${}^{-1}$ (M${}_2$, at 0 GPa) appear between $T_{N,1}$ and $T_{N,2}$ \cite{Song2022,Liu2020}. These spectroscopic features provide a robust signature of the magnetic phase transitions.

\begin{figure*}[htb!]
\centering

\includegraphics[width = 18.0cm]{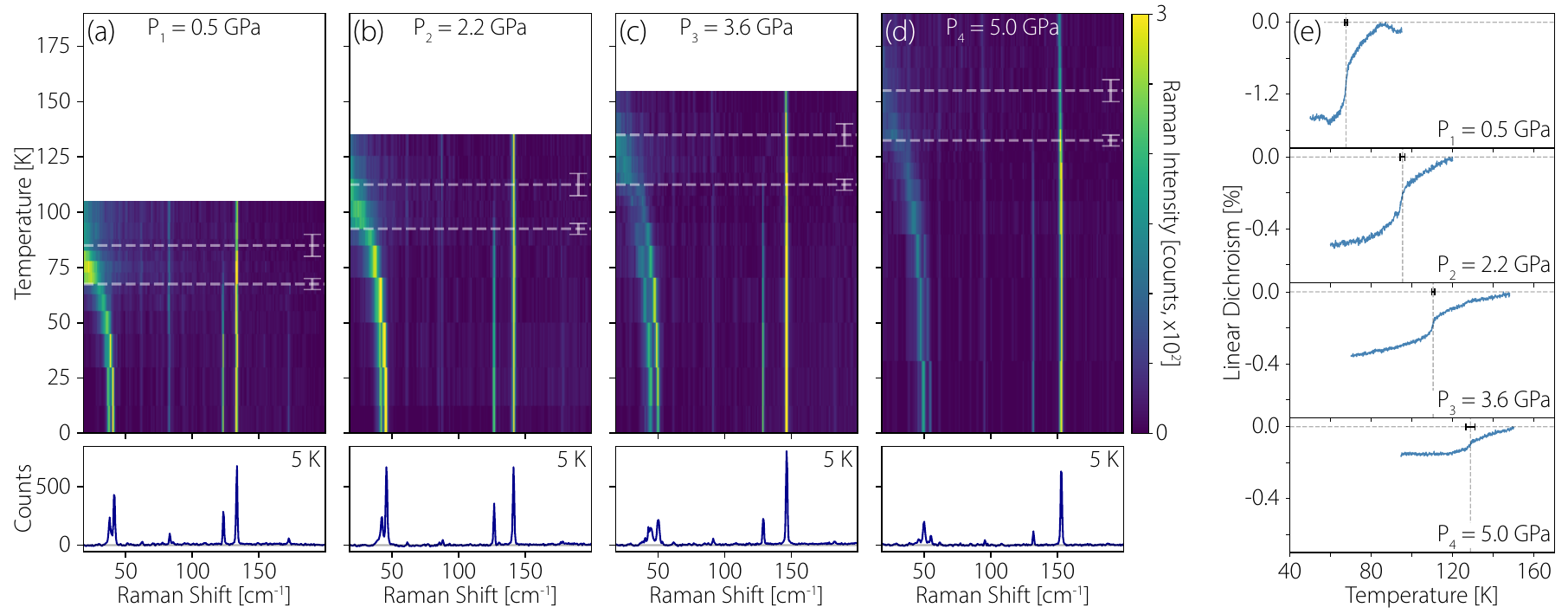}

\caption{(a)-(d), Temperature-dependent Raman maps for pressures $P_1$-$P_4$ (top) with the base temperature (5 K) spectra (bottom). The $T_{N,1}$/$T_{N,2}$ transitions are identified with the upper/lower horizontal dashed lines, respectively, with error bars indicated. (e), Temperature-dependent LD for $P_1$-$P_4$. The identified $T_{N,2}$ transitions are marked by vertical dashed lines, with error bars determined by the transition width. Note: the LD scale for the $P_1$ data (recorded with $\lambda = 633$ nm) is different from the $P_2$-$P_4$ data ($\lambda = 650$ nm). The LD curves are raw data shifted by a constant offset. }

\label{fig:fig2}
\end{figure*}

Figures \ref{fig:fig2}(a)-(d) show corresponding temperature-dependent XY-polarized Raman maps with increasing pressure between $P_1 = 0.5$ GPa and $P_4 = 5.0$ GPa, acquired with laser wavelength $\lambda = 532$ nm. At all measured pressures, the spectra are qualitatively similar to the results at ambient pressure \cite{Song2022,Liu2020}, except for a pressure-induced hardening of all modes, a change in the relative intensity of the modes and, importantly, a clear increase of both the $T_{N,1}$ and $T_{N,2}$ transition temperatures. The linear evolution of the Raman mode frequencies up to $5$ GPa and their spatial homogeneity across the sample at $5$ GPa confirm quasi-hydrostatic pressure conditions without significant deviatoric stresses \cite{SuppRef,Martins2023}. We estimate the resulting transition temperatures using the onset of the EM excitations as shown in Fig. \ref{fig:fig1}(e). The transition points are indicated by dashed white lines for each pressure, and are further supported by the temperature-dependent intensity of the $M_1$ peaks which are clearly shown to activate between $T_{N,1}$ and $T_{N,2}$ in Fig. \ref{fig:fig2}(a)-(d) (for detailed analysis see Ref. \cite{SuppRef}). Ambient pressure measurements on the recovered sample after decompression show the same transition temperatures as reported previously \cite{Song2022, SuppRef}. 

From these results, we immediately observe a clear increase of both the $T_{N,1}$ and the $T_{N,2}$ transition temperatures from $75$/$60$ K at ambient pressure \cite{Song2022} to $155 \pm 5$/ $132.5 \pm 2.5$ K at $P_4 = 5.0$ GPa, suggesting that both the AFM and the helimagnetic/multiferroic states are rendered more stable upon increasing pressure. To verify this, we also performed temperature-dependent LD measurements as a function of pressure (Fig. \ref{fig:fig2}(e)) in order to confirm the breaking of the three-fold rotational symmetry by the single-{\bf Q} helical state/ferroelectric order. These measurements were performed at the same pressure conditions as the corresponding Raman data, on a separate thermal cycle. At all pressures ($P_1$-$P_4$), the appearance of a LD signature indicates a first-order transition at increasing temperatures as pressure is increased. We extract the corresponding helimagnetic transition temperatures indicated in Fig. \ref{fig:fig2}(e), which agree with the independently-determined $T_{N,2}$ values from Raman.

We observe that the optical signatures of multiferroic order from Raman and LD appear to decrease in magnitude at high pressures despite the enhanced transition temperatures. Indeed, on a second experimental run, the multiferroic soft modes are absent at base temperature $T = 5$ K and pressure $P = 7.5$ GPa under otherwise identical experimental conditions \cite{SuppRef}. We attribute the decrease in magnitude of these optical signatures to a reduction of the charge transfer gap \cite{Pasternak1990, Chen1993, theoryref} which then futher detunes the optical transitions away from the excitation wavelengths. Thus, to provide additional evidence for a pressure-stabilized multiferroic transition at pressures above $5$ GPa and to confirm the persistence of magneto-structural coupling, we performed high-pressure, temperature-dependent XRD.

%Fig 3
\begin{figure}[htb!]
\centering

\includegraphics[width = \columnwidth]{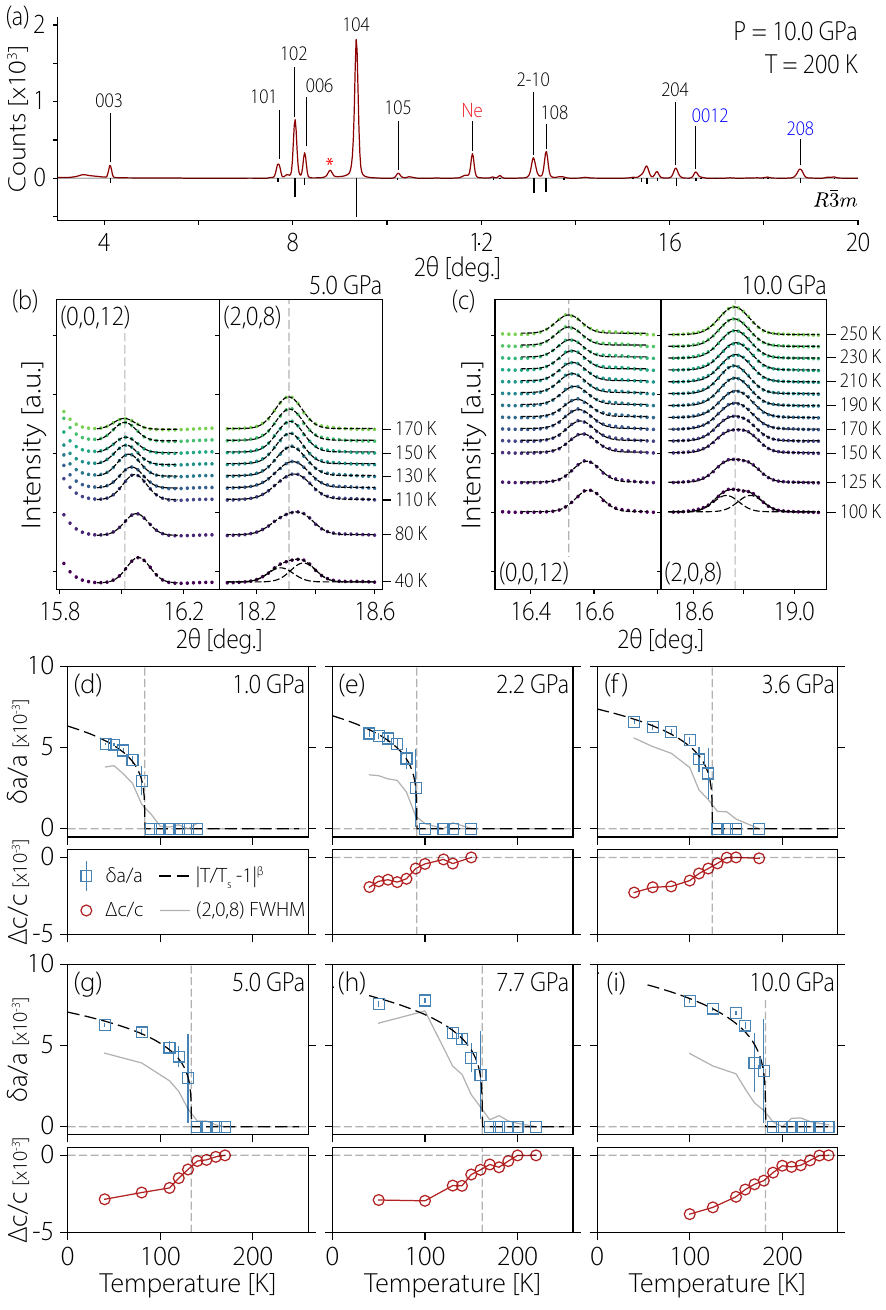}

\caption{(a) XRD spectrum recorded at $10$ GPa and $200$ K, indexed on the $R\bar{3}m$ structure (black bars). Temperature-dependent XRD of the $(0,0,12)$ and $(2,0,8)$ reflections at (b), 5.0 GPa and (c), 10.0 GPa. Gaussian peak fits are shown as dashed black lines. (d)-(i), Temperature-dependent splitting of the $a$ lattice parameter $\delta a/a$ (top) and the change in $c$ lattice parameter $\Delta c/c$ (bottom) for pressures between 1.0-10.0 GPa. The FWHM of the $(2,0,8)$ peak from a single-peak fit is shown as a grey line overlaid on the $\delta a/a$ plots. Black dashed lines are order parameter fits of the form $|T/T_S -1|^\beta$, with $\beta = 0.23$ as determined from the simultaneous best fit value for all pressures. The identified transitions ($T_S$) for each pressure are marked by vertical dashed grey lines.}

\label{fig:fig3}
\end{figure}

%Fig 4
\begin{figure}[htb!]
\centering

\includegraphics[width = \columnwidth]{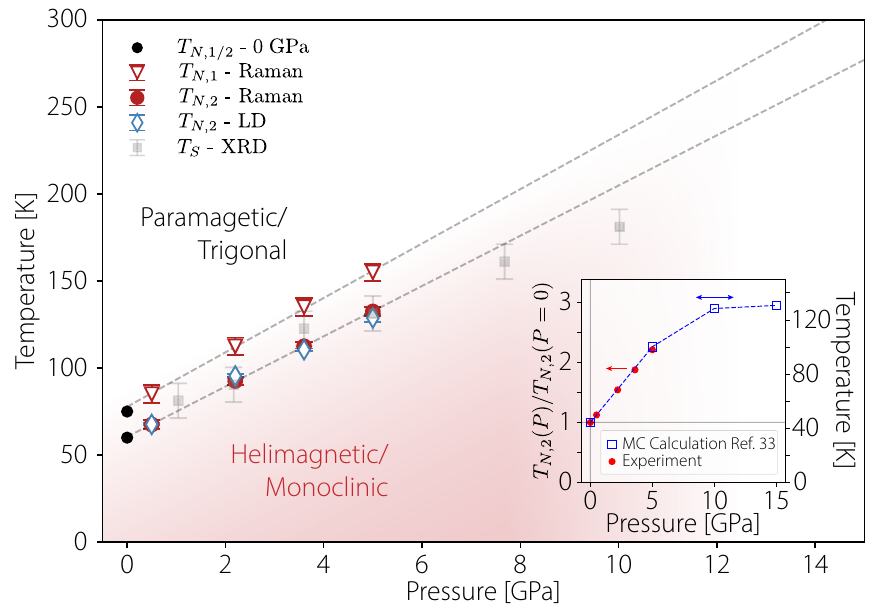}

\caption{Experimental phase diagram compiled from Raman, LD and XRD data. Ambient pressure data are from Ref. \onlinecite{Song2022}. Inset: helimagnetic transition temperature from first principles and Monte Carlo calculations (blue, left) from Ref. \cite{theoryref}. Overlaid are the experimental $T_{N,2}$ values from Raman, normalized by the ambient pressure value for comparison (red, right).}

\label{fig:fig4}
\end{figure}

XRD measurements were performed at Sector 16-ID-B of the Advanced Photon Source, Argonne National Laboratory, using an incident energy $E = 29.2$ keV, neon pressure transmitting medium and high-quality NiI$_2$ powder grown by chemical vapor transport. Fig. \ref{fig:fig3}(a) shows a representative $2\theta$-scan of the NiI$_2$ powder at $T = 200$ K and $P = 10.0$ GPa, indexed to the $R\bar{3}m$ high-temperature structure \cite{Kuindersma1981}. For all pressures considered (up to 10.0 GPa), the high-temperature structure remains trigonal \cite{SuppRef, theoryref}. Temperature- and pressure-dependent diffraction scans were acquired by measuring the diffraction with monotonically-decreasing temperature from above the transition to the base temperature $\simeq 40$ K. The pressure at a given thermal cycle was carefully stabilized using an {\it in-situ} double-sided membrane-driven DAC and monitored via ruby fluorescence.

To analyze the magneto-structural transitions, we focus on the $(2,0,8)$ and the $(0,0,12)$ reflections ($(h,k,l)$ in reciprocal lattice units in the high-temperature $R\bar{3}m$ unit cell, shown in blue in Fig. \ref{fig:fig3}(a)). Figs. \ref{fig:fig3}(b)/(c) show the temperature-dependence of the $(2,0,8)$/$(0,0,12)$ peaks at two representative pressures $P = 5.0$ GPa and $P = 10.0$ GPa, respectively. At each pressure, we observe a sharp change in the $(0,0,12)$ $2\theta$ position which is nearly concomitant with the splitting of the in-plane $(2,0,8)$ peak. The emergence of AFM order naturally leads to magnetostriction along the $c$-axis, which is observed in isostructural NiBr$_2$ \cite{Nasser1992} and consistent with our observations. Meanwhile, the splitting of the $(2,0,8)$ reflection corresponds to reduced rotational symmetry, consistent with the conclusions of previous ambient pressure diffraction studies \cite{Kuindersma1981} and those reached from optical LD measurements (Fig. \ref{fig:fig2}(e)). 

We extract the corresponding pressure- and temperature-dependent lattice parameters to determine the splitting of the in-plane lattice parameter ($\delta a/a$, Figs \ref{fig:fig3}(d)-(i), top) and the change in $c$-axis with respect to temperature ($\Delta c/c_0$, Figs. \ref{fig:fig3}(d)-(i), bottom). The onset of these anomalies occur at higher temperatures with increasing pressure. The relatively coarse temperature step-size precludes a definitive association of these structural signatures to two distinct magnetic phases, but are used here as independent evidence of magnetically-induced structural modulation. We thus estimate the corresponding transition temperatures ($T_S$) by fitting the $\delta a/a$ data to an order parameter function ($|T/T_s -1|^\beta$, Fig. \ref{fig:fig3}(d)-(i)), revealing a pressure-stabilized structural transition.

Overall, the smooth evolution of the Raman, LD and XRD data suggest the persistence of the AFM and helimagnetic/multiferroic phases up to 10 GPa with a nearly three-fold enhancement of the transition temperatures. Furthermore, there is no evidence for additional pressure-induced phases, from the absence of e.g. new Raman-active modes or diffraction peaks. The results of all measurements are summarized in the magneto-structural phase diagram of Fig. \ref{fig:fig4}. In the low pressure regime, a nearly linear increase of both the AFM ($T_{N,1}$) and the multiferroic ($T_{N,2}$) transitions is observed, with rates of $15.3$ and $14.4$ K/GPa, respectively, with strong agreement among all techniques. We note that these pressure-enhancements agree with the previous estimations from M\"{o}ssbauer spectroscopy \cite{Pasternak1990} and are on the same order of the observed rates of pressure-induced changes in the low-dimensional multiferroics CuBr$_2$ ($\sim 19.6$ K/GPa) \cite{Zhang2020} and NiBr$_2$ ($\sim -22.5$ K/GPa) \cite{Adam1981} while being stronger than typical 3D materials (e.g. CuO $\sim 3.5$ K/GPa \cite{Terada2022}). This suggests that the strong tunability of magnetic order in low-dimensional materials with pressure is a general approach for increasing $T_N$ in type-II multiferroics.

In order to interpret the mechanisms behind the increased magnetic phase transition temperatures, we consider the pressure-dependence of the magnetic exchange interactions which were recently determined from first-principles calculations \cite{theoryref}. These reveal that while $J^{\parallel}_1$ is weakly pressure-dependent, $J^{\parallel}_3$ and $J^\perp_\text{eff}$ are strongly enhanced with increasing pressure \cite{theoryref}. The associated transition temperatures determined from Monte Carlo calculations in Ref. \onlinecite{theoryref} are shown in comparison to the experimentally-determined values in Fig. \ref{fig:fig4} (inset). While there are small quantitative discrepancies, normalizing the calculated and experimental $T_{N,2}$ temperatures by their corresponding ambient pressure values shows that the relative enhancement is well-captured up to 5 GPa. The saturation of $T_{N,2}$ above 10 GPa in the calculations likely indicates a competitive effect of $J^\perp_\text{eff}$ and $J^\parallel_3$ which have a distinct dependence on pressure \cite{theoryref}. While a small deviation from linearity of $T_S$ is observed above 5 GPa, investigations extended to higher pressures will be needed to confirm this scenario. 

The pressure-enhancement of the ratio $\big|J^{\parallel}_3/J^{\parallel}_1\big|$ has important implications for the helimagnetic propagation vector, which can be analytically determined (assuming $J_\text{eff}^\perp = J_2^\parallel = 0$) as \cite{Kuindersma1981, theoryref, Rastelli1979} 
\begin{equation}
\mathbf{Q}_\parallel = 2 \arccos \left[\left(1 + \sqrt{1 - 2J_1^\parallel/J_3^\parallel}\right)/4\right]
\end{equation}
This, in turn, impacts the spin-induced ferroelectric order which can be estimated by the generalized Katsura-Nagaosa-Balatsky (gKNB) model as $\mathbf{P} \propto \sin \mathbf{Q}_\parallel$ \cite{Katsura2007, Song2022, theoryref, Tokura2014, Kurumaji2013, Xiang2011}. Thus, the observed increase in $\big|J^{\parallel}_3/J^{\parallel}_1\big|$ with pressure favors a shorter in-plane spiral pitch (larger $\mathbf{Q}_\parallel$) with larger nearest-neighbor spin angle and, potentially, a larger spin-induced polarization \cite{Tokura2014}. It will be insightful in future studies to directly measure the predicted change of $\mathbf{Q}_{\parallel}$ with neutron scattering, and the modifications to the electrical polarization $\mathbf{P}$ with SHG or pyroelectric current measurements. 

In conclusion, we have demonstrated a significant enhancement of the helimagnetic order in bulk NiI$_2$ with hydrostatic pressure. Combined with first-principle calculations, this reveals the central role of interlayer and third-nearest neighbor exchange interactions for stabilizing multiferroic order in an archetypal vdW multiferroic. It will be interesting to investigate the behavior of these pressure-stabilized magnetic/multiferroic orders near the charge transfer gap closure, where the charge-transfer energy becomes comparable to other excitations of the material \cite{Chen1993}, possibly leading to novel quantum phases. Finally, an explicit demonstration of magnetoelectric switching would pave the way for the wider use of NiI$_2$ in devices.

{\it Acknowledgements.}
C.A.O. and L.G.P.M. contributed equally to this work. C.A.O., L.G.P.M., Q.S., and R.C. acknowledge support from the US Department of Energy, BES under Award No. DE-SC0019126 (materials synthesis and characterization, optical measurements, and X-ray diffraction measurements). D.A., B.D. and M.J.V. acknowledge the SWIPE project funded by FNRS Belgium grant PINT-MULTI
R.8013.20. M.J.V. acknowledges ARC project DREAMS (G.A. 21/25-11) funded by Federation Wallonie Bruxelles and ULiege. J.K. and A.S.B. acknowledge support from NSF Grant No. DMR 2206987 and the ASU Research Computing Center for high-performance computing resources.


%\bibliography{NiI2_HighPressure_Bib_20230612}

\end{document}